\documentclass[modern]{aastex7}
\usepackage{multirow}
\usepackage{adjustbox}
\usepackage{tabularx}
\usepackage{adjustbox}
\usepackage{float}
\usepackage{caption}
\usepackage{subcaption}

\begin{document}

\title{Candidate Captured Interstellar Objects in the Solar System}

\author[orcid=0009-0003-2157-7036]{Rainer Marquardt-Demen}
\affiliation{Georgia Institute of Technology, School of Physics}
\email[show]{rmarquar3@gatech.edu}  

\author[orcid=0000-0001-8308-0808, gname='Gongjie', 
sname='Li']{Gongjie Li} 
\affiliation{Georgia Institute of Technology, School of Physics}
\email[show]{gongjie.li@physics.gatech.edu}

\author[orcid=0000-0001-5559-2179]{James J. Wray} 
\affiliation{Georgia Institute of Technology, School of Earth and Atmospheric Sciences}
\email[show]{jwray@gatech.edu}
 
\begin{abstract}
Interstellar objects (ISOs) provide direct probes of planetesimal formation and ejection in other planetary systems. While most ISOs that pass through the Solar System escape it after a single passage, a small fraction can become temporarily bound through gravitational interactions. We develop a self-consistent semi-analytic framework that couples analytic modeling of the interstellar object flux near Jupiter with N-body simulations of capture and long-term dynamical evolution, allowing us to predict the steady-state phase-space distribution of bound interstellar objects. Using an analytic model to construct initial conditions and N-body integrations to simulate capture and ejection, we compute capture rates and orbital distributions consistent with previous analytical estimates, finding a mean capture interval of approximately 220 years. We show that post-capture survival strongly reshapes the observable population: although capture initially favors prograde orbits, long-term stability is dominated by highly inclined objects that encounter planets less frequently and thus are less likely to be ejected. The resulting steady-state population is therefore concentrated at high inclinations, providing a potential dynamical discriminant for identifying candidates. However, most captured objects occupy semimajor axes comparable to the inner Oort cloud, making them difficult to distinguish from native long-period comets. The predicted phase-space distribution contains 122 known Solar System objects within the highest-density region of our model. 

\end{abstract}


\section{Introduction} \label{sec:background}

Interstellar objects (ISOs) are planetesimals no longer gravitationally bound to a star \citep{Seligman_2022, Siraj_2022}. ISOs passing through our Solar System are called interstellar interlopers \citep{Jewitt_2023}. Only a few interstellar interlopers have been detected; however, the discoveries of 1I/'Oumuamua (2017), 2I/Borisov (2019), and 3I/ATLAS (2025) sparked active pursuit in the astronomical community \citep{Siraj_2022, Seligman_2022, Jewitt_2024, Jewitt_2023, Fitzsimmons_2024, Seligman_2025, de_la_Fuente_Marcos_2025}. Improved observations can constrain galactic composition and reveal the physical mechanisms ejecting small bodies from parent systems. Earlier models of planetesimal composition and ejection depended solely on the Solar System. Interlopers can broaden these models and expand the search for evidence of life beyond Earth.

To access this frontier of astronomy, we require larger sample sizes and a broader approach to studying these objects. Upcoming surveys, including the Rubin Observatory Legacy Survey of Space and Time (LSST) and the Near-Earth Object Surveyor mission, are expected to significantly increase the detection rate of interstellar interlopers \citep{Jewitt_2023}. Ideally, we would send probes to intercept an interloper and collect in situ data \citep{Seligman_2018, Moore_2021, Stern_2024}. However, the expected high speeds and low number density of interlopers make planning such a mission extremely difficult. Instead, we consider cases where the interloper does not escape. Although rare, the combined gravitational influence of large planets like Jupiter and our Sun can perturb interlopers, causing them to lose energy and remain (at least temporarily) bound to our Solar System \citep{Dehnen_2021}. 

Given the possibility of bound ISOs, we ask how to identify or constrain which Solar System objects may be of interstellar origin. In this work, we develop a self-consistent semi-analytical framework to model the capture, long-term stability, and orbital distributions of captured ISOs, enabling direct comparison with known Solar System populations. We expect that captured ISOs may exhibit distinctive orbital properties due to the strict dynamical criteria required for their capture. The only other population with comparable dynamics would be Oort Cloud objects that have migrated inward, as these objects are approximately isotropic around the Sun \citep{Hands_2020, Siraj_2019}. We find that capture and survival probabilities depend sensitively on orbital inclination, introducing a strong observational discriminant that has not been explored in previous studies. From future observations of interstellar interlopers, additional characteristics, such as non-gravitational acceleration and composition, can further constrain whether an object is of interstellar origin.

Several previous studies have investigated the capture process through simulations, with varying results \citep{Dehnen_2021, Napier_2021a, Napier_2021b, Siraj_2019, Hands_2020, Mukherjee_2023}. However, only one, \citet{Siraj_2019}, directly compared simulated capture outcomes with known Solar System populations to identify captured ISO candidates. That study relied on a simplified distribution of ISOs near Jupiter and did not account for post-capture dynamical stability, both of which significantly affect the inferred capture rate and candidate population. Rather than reanalyzing prior simulations, we reconstruct the capture calculation from first principles by analytically deriving the ISO phase-space distribution near Jupiter and coupling it to N-body integrations that track capture, stability, and ejection. \citet{Dehnen_2021} also studied the capture of the ISO population in the Solar System; however, they estimated the lifetimes and ejections of these ISOs only analytically, and did not consider inclination in their distributions. In this work, we generalize this by simulating each ISO's lifetime from capture to potential ejection, and analyzing the orbital distributions of captured ISOs.

The paper is as follows: First, we describe the N-body simulations used to investigate the capture rate and orbital element distributions of bound ISOs. This is divided into three parts: checking the isotropic distribution after scattering by the Sun, Monte Carlo N-body capture simulations, and long-period simulations of ejection. Then we present our results and compare them to the analytical results of \citet{Dehnen_2021}. We also compare the resulting steady-state population to known objects in the Solar System. Finally, we discuss the significance of our results.

\section{Simulation Framework} \label{sec:N-Body}

The general set-up of these simulations is as follows: We first initialize a population of ISOs in a shell of radius $R=250$ au centered on Jupiter's instantaneous position (evaluated at mean orbital elements). We assume that only Jupiter, Saturn, and the Sun contribute meaningfully to the capture of ISOs, as previous work suggests \citep{Dehnen_2021, Siraj_2019}. Although we assume that the distribution of ISOs at infinity is uniform and isotropic, in general, the distribution of ISOs on this shell is not. Instead, the distribution is gravitationally focused by the Sun (and to a lesser extent, by Jupiter and Saturn). For ease of computation, we first analytically compute the hyperbolic scattering due to the Sun alone. We simulate forward until the ISO has escaped, become bound, or experienced a collision. For those that have become bound, we continue simulating until nearly all captured ISOs are ejected and save the object's orbital parameters over its lifetime. We adopt the REBOUND N-body package for the N-body simulations \citep{Rein_2012, Tamayo_2019}. 

\subsection{Initial Distribution} \label{subsec:iiso}

Each ISO is modeled as a massless point particle. We assume the distribution of ISOs in both position and velocity is isotropic at infinity. We sample impact parameters $b$ with respect to the Sun uniformly in $b^2$, which corresponds to an isotropic flux at infinity. The velocity distribution of ISOs at infinity is assumed to follow the local stellar velocity distribution. We approximate this with a 3D Maxwell-Boltzmann distribution. As all objects are assumed to have a non-zero speed at infinity, they start out unbound to the Solar System.

For a given impact parameter $b$, initial velocity at infinity, and azimuthal angle, we can solve the hyperbolic scattering (gravitational focusing) from only the Sun analytically and check that the object enters a sphere of radius $R$ around Jupiter, and determine the object's position and velocity at entry into the sphere. ISOs that never enter this sphere are discarded, since they are unlikely to become bound. This process reduces simulation time, but also incurs error due to the presence of Jupiter and Saturn, and is highly dependent on the value of $R$. By testing increasing values of $R$, we find convergence at $R\approx250$ au.

We sample up to some maximum impact parameter defined by:

\begin{equation}
    b_{max} = (a_J + R)\sqrt{1 + \frac{2GM_\odot}{(a_J + R)\cdot(v_{min}^2)}}
\end{equation}

Here, $a_J$ is the semimajor axis of Jupiter and $v_{min}$ is the minimum speed sampled. Beyond this impact parameter, no ISOs will be able to enter the sphere $R$. For computational reasons, we also set maximum limits on sampled velocity and eccentricity for the initial trajectory.  We do not sample velocities higher than $0.8$ au/yr, as previous studies show capture of objects with this initial velocity is unlikely \citep{Siraj_2019, Dehnen_2021}. We find that capture from initially highly eccentric hyperbolic orbits with $e>1.3$ is also unlikely.

\subsection{Monte Carlo Capture Simulation}\label{subsec:mc}

Now that we have our initial distribution, that being the velocities and positions of ISOs located on the sphere of radius $R$ found above, we integrate up to a maximum time $T=10$ million years. After this time, nearly all particles have collided, escaped, or become captured. After each integration step $\Delta t=10$ years, we check whether the particle has become bound to the Sun ($e<1$), collided, or escaped (distance from the Jupiter-centered sphere exceeds $R$) during the integration. 

For each particle that becomes bound, we save its initial parameters, its orbital parameters immediately after becoming bound, and the time of capture. To calculate the capture rate, we bin the captured objects by velocity and calculate the capture cross-section for each bin, defined by:

\begin{equation}
    \sigma(v) = \pi b_{max}^2\frac{N_{captured}(v)}{N_{total}(v)}
\end{equation}

Where the bins are centered at $v$. Then the capture rate is given by:

\begin{equation}
    \Gamma = n_{iso}\int vf(v)\sigma(v)dv
\end{equation}

Where $n_{iso}$ is the spatial density of ISOs at infinity, taken to be $0.1$au$^{-3}$,  consistent with estimates from recent ISO detections, and $f(v)$ is the probability density function for our 3-D Maxwell velocity distribution \citep{Dehnen_2021, Siraj_2019, Do_2018, Eubanks_2021}. We again bin with respect to the semimajor axis of the bound ISO right after capture to find the differential capture rate $\frac{d\Gamma}{da}$. 

\subsection{Lifetime Simulation} \label{subsec:life}

Next, we investigate the long-term stability of the captured ISOs. Using the initial parameters saved from the bound ISOs, we initialize a new simulation with the captured ISO, Jupiter, Saturn, and the Sun, using the orbital elements of the bound ISO right after capture. We integrate each captured object forward for up to $10^6$ times the mean capture interval, corresponding to a maximum integration time of $\sim 2 \cdot 10^8$ years, by which point nearly all captured objects have been ejected. On each timestep $\Delta t=10$ years, we save the orbital parameters and check whether the particle has collided or escaped using the same criterion described for the capture simulation. If the particle collides or escapes during that timestep, we record its lifetime. If not, we continue simulating. Then, the differential steady state population is given by:

\begin{equation}\label{eq:steady}
    \frac{dN(a,q,i)}{da}=T(a, q, i)\frac{d\Gamma}{da}
\end{equation}

Where $T(a,q,i)$ is the average time from capture to ejection for ISOs with semimajor axis $a$, perihelion $q$, and inclination $i$, from this, we see that even if capture is rare for certain orbits, these objects can dominate the steady-state population if they have a large mean residence time. Note that $T$ is not only dependent on $a$, but also has dependence on inclination and eccentricity as well. To account for this, we construct the phase-space density in terms of the number of objects within small bins in the three-dimensional space of semimajor axis, inclination, and perihelion. Since the expected eccentricities are very close to one, it is easier to compare using perihelion, which is dependent on eccentricity. For each 3-D bin, we weight the phase-space density not just on the number of objects in each bin, but also on the average time until ejection for objects in the bin. This determines the steady-state phase-space density, which we use to determine the highest density regions (HDR) of orbital-phase space that observers would expect to see bound ISOs in. Using a conservative $90\%$ HDR, the smallest region of phase space containing $90\%$ of simulated bound objects, we then search through the JPL small-body database for known Solar System objects and label them as possible candidates of bound ISOs. 

\pagebreak
\section{Results} \label{sec:results}

\begin{figure}
    \centering
    \includegraphics[width=\linewidth]{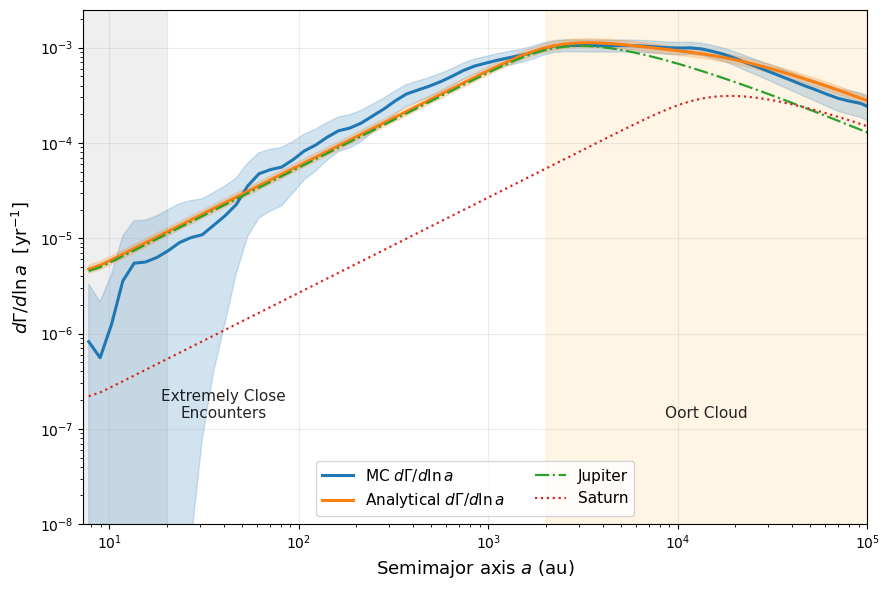}
    \caption{Differential capture rates for Monte Carlo (MC) simulation results compared with the analytical estimates from \citet{Dehnen_2021} (their Equations 3 and 13). The dashed lines correspond to the individual contributions from Jupiter and Saturn, respectively, to the total analytical differential rate. The analytical shaded region represents a $10\%$ uncertainty, and the Monte Carlo shaded region represents a $3\sigma$ uncertainty of the estimator. The capture rate contributes nonnegligibly past $10^5$ au, but is negligible observationally.}
    \label{fig:mcestimate}
\end{figure}

After simulating about 380 million ISO trajectories, we found that 15000 ISOs became captured. Using the Monte Carlo estimator described in Section \ref{subsec:mc}, we find that the mean time between ISO captures is $\sim220$ years. This matches well with the analytical results of \citet{Dehnen_2021}, as seen in Figure \ref{fig:mcestimate}.  

In Figure \ref{fig:mcestimate}, we find good agreement between the differential capture rate from previous literature and our own results in regions where the capture rate is high. In particular, the uncertainty of each bin (the shaded region), set at $3\sigma$, overlaps with the fitting function for Jupiter and Saturn for all semimajor axis bins. Note that the uncertainty spikes in regions of small capture probability, where the small number of objects in each bin dominates. These regions also correspond to areas where the analytical estimates of \citet{Dehnen_2021} become inaccurate. Especially for close encounters ($a<2a_J$), shaded in gray in Figure \ref{fig:mcestimate}, collisions become non-negligible, which the analytical estimates do not include.

These results suggest that most objects are captured onto orbits with a semimajor axis corresponding to the inner edge of the Oort cloud, making them difficult to observe and distinguish from Solar System objects. However, the semimajor axis is only one phase-space dimension. 

\subsection{Orbital Element Distribution of Captured Interstellar Objects}\label{subsec:orbit}

\begin{figure}[h]
    \begin{subfigure}{.5\textwidth}
        \includegraphics[width=1\linewidth]{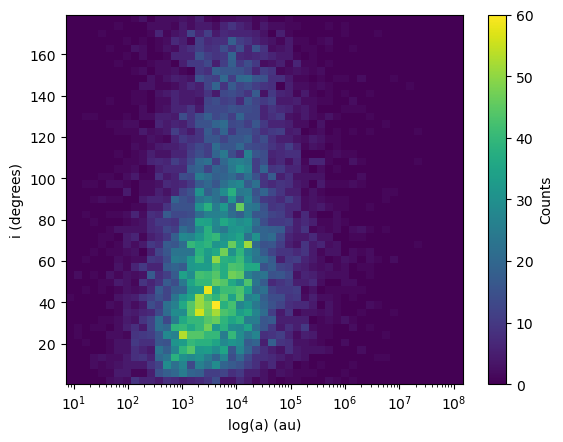}
        \caption{Inclination vs. Semimajor Axis}
    \end{subfigure}
    \begin{subfigure}{.5\textwidth}
        \includegraphics[width=1\linewidth]{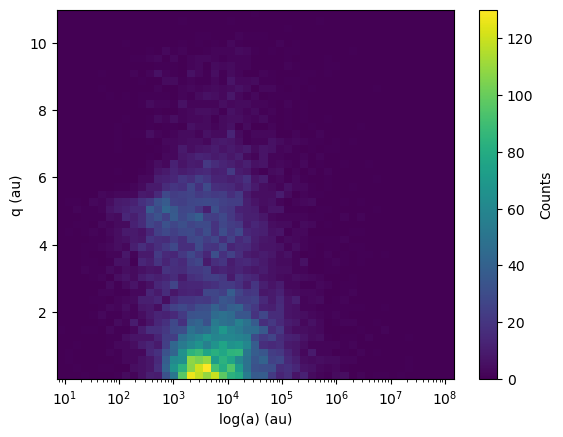}
        \caption{Perihelion vs. Semimajor Axis}
    \end{subfigure}
    \caption{Orbital element distributions immediately after capture.}
    \label{fig:idist}
\end{figure}

Immediately after capture, the distribution of captured ISOs shows a preference for prograde orbits and extremely small perihelion distances, as shown in Figure \ref{fig:idist}. 

For an isotropic population, the inclination distribution should follow a $\sin i$ distribution. Instead, the capture mechanism prefers certain inclinations. There are two main capture mechanisms: wide encounters, resulting from small perturbations of the Jupiter-Sun interaction, and close encounters, resulting from gravitational braking \citep{Dehnen_2022}. Wide encounters, which are the majority of all captures, have the maximum (negative) energy change for prograde encounters \citep{Dehnen_2022}. Since the deflections are small, the orbit the ISO is captured onto will also tend towards prograde. In contrast, capture from close encounters (slingshot) obtains the greatest (negative) energy change for retrograde encounters. Since wide-encounters have a much larger capture cross-section, the total distribution is roughly dominated by prograde directions and the assumption of isotropy at infinity, resulting in a peak at around $i=40^\circ$.

Similarly, we see that the perihelion is influenced by the type of capture. Here, the distribution is bimodal, where wide encounters in general have extremely small perihelion distances, on the order of $q<2$ au. Close encounters instead correspond to greater energy changes and are more closely bound with perihelion of the order $q\sim5$ au. 

In general, these distributions are not the same as the orbital element distributions for the steady state population. Any specific captured ISO will rarely have a drastic change to its orbital elements that keeps the ISO bound. Most perturbations after capture rather lead to ejection. However, certain orbital elements may be more favored for ejection than others. 

\begin{figure}
    \centering
    \includegraphics[width=\linewidth]{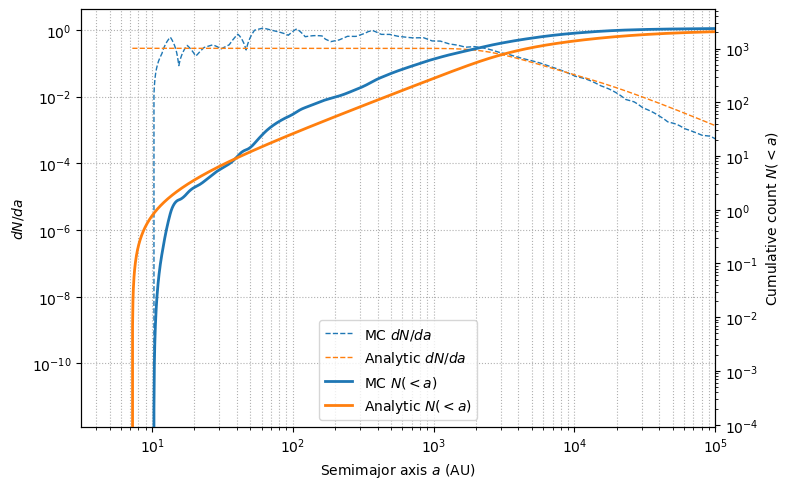}
    \caption{Differential and cumulative steady-state population $N$ with respect to semimajor axis. The analytical portion in orange is obtained by multiplying the previous analytical differential capture rate in Figure \ref{fig:mcestimate} by the mean residence time from our simulations. Note that the mean time past $\sim10^5$ au (beyond the Oort cloud) becomes inaccurate and an overestimate, as galactic tides and passing stars become important.}
    \label{fig:steadypop}
\end{figure}

From our simulations in Section \ref{subsec:life}, the mean ejection time is on the order of $\sim$Myr, implying that there should be a non-zero steady-state population of bound ISOs in our Solar System (see Figure \ref{fig:steadypop}). This steady-state population is lower than the phase-space estimates in \citet{Dehnen_2021}, which predict $\sim 2000$ bound ISOs between $4$ and $2000$ au. In particular, they also predict several bound ISOs residing in bound orbits with $a < 10$ au at any time. For the smaller semimajor axes, we find this to be an overestimate. The phase-space-based estimate assumed that the semimajor axis dominated the phase-space volume and neglected collisions, which is important for closely bound ISOs. Numerical simulations of ejection are more accurate in this case, except for extremely wide ISOs, where factors not included in our numerical simulations, such as galactic tides and passing stars, are important. For these wide ISOs, the phase-space estimate becomes more accurate; however, these ISOs are typically short-lived and do not contribute strongly to the steady-state population.

\begin{figure}[h]
    \centering
    \includegraphics[width=0.9\linewidth]{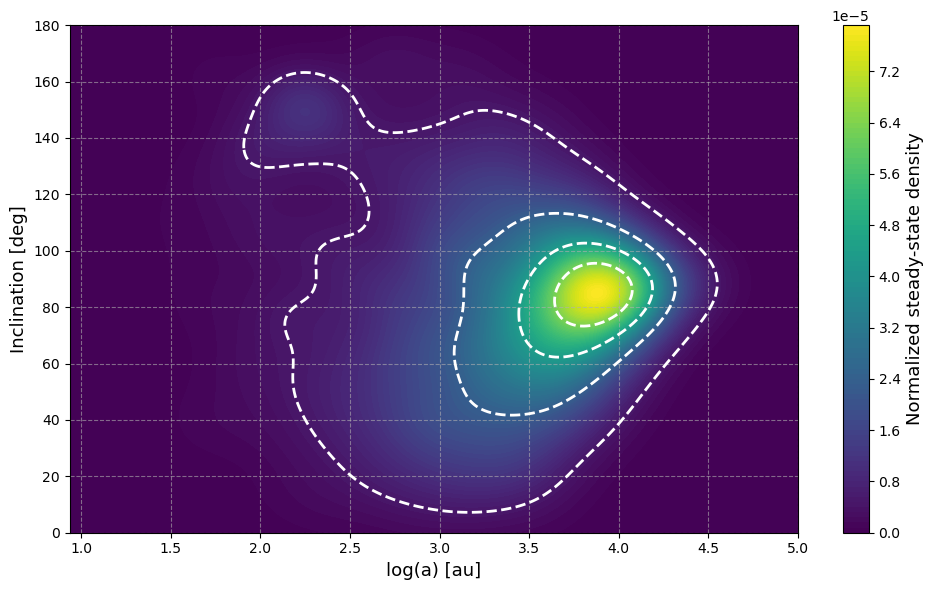}
    \caption{Normalized steady-state orbital phase-space density. The white contours represent the highest density regions of $10\%$, $25\%$, $50\%$, and $90\%$, where the $90\%$ region corresponds to the outermost contour.}
    \label{fig:phasespace}
\end{figure}

From equation \ref{eq:steady}, the steady-state population orbital phase-space is biased by the residence time. Thus, we are more likely to observe more stable ISOs. We find that residence time is only weakly affected by the perihelion and semimajor axis, but very strongly affected by inclination. Objects with highly inclined orbits $i\sim90^\circ$ have much higher residence times and dominate the total steady-state population. This can be seen in Figure \ref{fig:phasespace}, where we plot the density of $N(a,q, i)=T(a,q, i)\Gamma(a,q, i)$ per bin in orbital phase-space.

The reason that highly inclined captured ISOs are more stable is that the ejection mechanism is often the same as the capture mechanism: encounters with a massive planet. Although this encounter may be wide or close, it still requires the planet to exert its gravity non-negligibly on the captured ISO. For highly inclined orbits, the ISO is much less likely geometrically to encounter a planet on a prograde orbit in such a way that causes non-negligible perturbations. Thus, even though capture is dominated by prograde inclinations, these orbits are much more likely to be ejected, and the steady-state population is dominated by highly inclined orbits ($i\sim 90^\circ$). 

\subsection{Comparison of ISOs to Solar System Objects}\label{subsec:cand}

Given the predicted steady-state orbital element distribution, we consider now how these objects compare to Solar System bodies. Many of the captured ISOs share similarities with long-period comets (LPCs), due to their high semimajor axis but low perihelion. Unfortunately, the majority of captured ISOs lie between $10^3$ and $10^5$ au, which corresponds to the Oort cloud. Since the Oort cloud is roughly isotropic and contains a much greater population of potentially active LPCs ($\sim 10^12$) than our estimated $~10^3$ captured ISOs, we cannot differentiate captured ISOs from Oort cloud objects via orbital elements only \citep{Wang_2014, Dehnen_2021, Boe_2019}. However, we can constrain where to look for possible captured ISO candidates using the HDR contours.

From our simulations, we compute the expected phase space volume of steady-state ISOs, and find that there exist 122 previously observed Solar System objects within the $90\%$ HDR contour in Figure \ref{fig:phasespace}. Given the expected captive ISO population compared to the size of the Oort cloud, it is therefore unlikely that any of these candidates can be confidently identified as captured ISOs. To determine whether an object was originally interstellar requires a discriminant of ISOs outside of orbital parameters, besides the obvious hyperbolic orbit ($e>1$). So far from data on known ISOs 1I/‘Oumuamua, 2I/Borisov, and 3I/ATLAS, there is no such discriminant that all objects share. Although 1I/‘Oumuamua in particular showed anomalous behavior such as non-gravitational acceleration without a coma and an unusual shape, 2I/Borisov and 3I/ATLAS exhibit regular cometary behavior (although they have slightly unusual compositions) \citep{Jewitt_2022, Seligman_2025, Cordiner_2025}. Future observations of interstellar interlopers, particularly in non-dynamical features such as composition, may help identify captured ISOs.

\section{Discussion}\label{sec:disc}

In this paper, we have constructed N-body simulations to determine the steady-state population of bound ISOs in our Solar System and compared the resulting orbital element distribution with known Solar System objects. By using a three-part simulation pipeline: analytical estimates for the initial conditions around Jupiter following scattering from infinity, N-body Monte Carlo simulations on the capture of ISOs, and long-term N-body simulations until ejection, we find that our results for capture match well with the literature. In particular, we find that the mean time between capture of ISOs is around $220$ years. However, we find modest deviations in the steady-state calculation. Our results yield a steady-state population of $\sim 2500$ captive ISOs throughout the entire Solar System. Considering collisions and long-term stability, this corresponds to a smaller steady-state captive ISO population than previously estimated by \citet{Siraj_2019} (estimated $\sim 4000$ bound ISOs in identifiable regions) or \citet{Dehnen_2021} (estimated $\sim 2000$ bound ISOs between $4$ au and $2000$ au). We also find that although capture prefers orbits with prograde inclinations, the steady-state population is dominated by highly inclined ($i\sim 90^\circ$) orbits, providing a new dynamical discriminant for identifying captured ISOs.

Most of these captured ISOs, however, are captured near the inner edge of the Oort cloud, making repeated observations difficult. It also makes identification as definitively interstellar unlikely, as dynamically captured steady-state ISOs have low perihelia and high inclinations, which are not distinct from most LPCs originating from the Oort cloud. We use the steady-state phase-space density to determine the most probable ($90\%$) region where captured ISOs reside, and find 122 possible candidates. Currently, there does not exist any discriminant that allows us to dynamically constrain this candidate list any further. However, future observations on detailed properties of these objects (e.g., their isotope composition---\citet{Cordiner_2026, Opitom_2026, SalazarManzano_2026}) may help draw the distinction. In the meantime, we conclude that interstellar interlopers with hyperbolic orbits offer the best method of studying ISOs.




\software{\href{https://github.com/rainermd/ISO-ISBOCapture}{GitHub Repository}, \href{https://ssd.jpl.nasa.gov/sb/}{JPL Horizons Database}, REBOUND \citep{Rein_2012}, REBOUNDx \citep{Tamayo_2019}, \href{https://www.dask.org/}{Dask}}

\bibliographystyle{aasjournalv7}
\bibliography{bibli}

\end{document}